# Analytic Deep Learning-based Surrogate Model for Operational Planning with Dynamic TTC Constraints

Gao Qiu, Youbo Liu, *Member, IEEE*, Junyong Liu, *Member, IEEE*, Junbo Zhao, *Senior Member, IEEE*, Lingfeng Wang, *Senior Member, IEEE*, Tingjian Liu, and Hongjun Gao, *Member, IEEE*

*Abstract*—The increased penetration of wind power introduces more operational changes of critical corridors and the traditional time-consuming transient stability constrained total transfer capability (TTC) operational planning is unable to meet the real-time monitoring need. This paper develops a more computationally efficient approach to address that challenge via the analytical deep learning-based surrogate model. The key idea is to resort to the deep learning for developing a computationally cheap surrogate model to replace the original time-consuming differential-algebraic constraints related to TTC. However, the deep learning-based surrogate model introduces implicit rules that are difficult to handle in the optimization process. To this end, we derive the *Jacobian* and *Hessian* matrices of the implicit surrogate models and finally transfer them into an analytical formulation that can be easily solved by the interior point method. Surrogate modeling and problem reformulation allow us to achieve significantly improved computational efficiency and the yielded solutions can be used for operational planning. Numerical results carried out on the modified IEEE 39-bus system demonstrate the effectiveness of the proposed method in dealing with complicated TTC constraints while balancing the computational efficiency and accuracy.

## I. INTRODUCTION

With the increased integration of renewable energy resources and flexible loads into today's modern power systems, the system is operated more and more close to their stability boundary. This requires the development of security control strategies to carefully monitor inter-area transfer limits. It can be achieved via the total transfer capability (TTC) monitoring and control tool [1], [2]. However, TTC is computationally expensive for the large-scale system as the complicated differential and algebraic constraints need to be assessed every time. To mitigate that, dispatchers usually develop an overly conservative empirical TTC value to operate systems, causing unnecessary limitations on transfers and not economically attractive solutions.

TTC assessment incorporates various security constraints (e.g., static security, transient stability) under numerous contingencies. The integration of wind generators and energy storage systems has further introduced more dynamic patterns into the systems, which aggravates the difficulty of TTC constrained operation planning (TCOP). Although there are several model-based methods for calculating stability constrained TTC, e.g., implicit integration rules [3]–[5], single machine equivalent [6], sensitivity analysis [7], [8], or a hybrid of the above methods [9], they are not efficient enough to be adopted for temporal TCOP. This is because it is quite time-consuming to assess differential-algebraic equations under multi-contingencies at every period, see Fig. 1 for example.

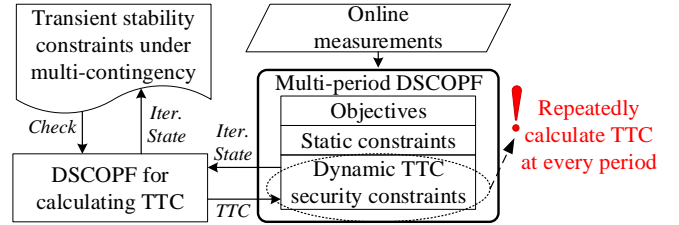

Fig. 1 The structure of the TCOP problem, where "DSCOPF" refers to dynamic security-constrained optimal power flow.

Recently, the surrogate-assisted method (SAM) has been advocated for fast decision-making of the problems involving sophisticated models and constraints [10]–[12]. SAM generally uses data-driven rules to surrogate the most computationally intensive model-driven parts in a decision-making task, which can significantly reduce problem complexity. Motivated by that, the concept of SAM is extended to solve the TCOP problem in this paper for the first time. With learning algorithms, such as decision trees [13]-[14], elastic net [15], and multiple regression [16], SAM has been developed for some power system applications. For example, multiple regression is used for TTC assessment [16]. However, oversimplification has been made, such as ignoring the operational costs, and time-varying characteristics of the TTC. It is worth noting that if the model nonlinearity is not appropriately addressed, large errors occur that has been validated in many security assessments or prediction applications [17]–[19]. Thus, to ensure the accuracy in both prediction and control, it is necessary to develop accurate and robust nonlinear surrogate models for the TCOP. However, the rules extracted by nonlinear learning algorithms are often in a "black-box" form (e.g., deep neural networks (DNNs), etc) and thus intractable for OPF. A pattern discovery method [20] is first proposed to identify the rules of transient trajectories, which are used later to replace the unstable operation condition with the nearest secure ones. This method is, however, unable to make an in-depth analysis of the rules. Han *et al.* propose a more systematic way to utilize implicit rules for dynamic VAR planning employing heuristic algorithms [21]. But, due to "the curse of dimensionality" issue, the existing gradient-free algorithms are unsuited for high-dimensional problems, such as the TCOP in this paper. To address the aforementioned challenges, this paper proposes a new surrogate model-based approach for the TCOP problem. It yields the following contributions:

- The deep learning algorithm is advocated to develop a computationally cheap surrogate model to replace the original time-consuming differential-algebraic constraints related to TTC. Both the fast time-varying factors (e.g., wind farms, energy storage systems, and unit ramp constraints) and dynam-ics of synchronous generators and wind generators are considered under various contingencies. This surrogate-assisted scheme allows us to significantly improve computational efficiency while capturing the dynamic TTC-defined boundary conditions. To this end, the surrogate-assisted TCOP (SA-TCOP) is formulated.

- The deep learning-based surrogate model introduces implicit rules/constraints in "black-box types" that are very challenging for the existing optimization algorithms to handle. To deal with that, we derive the *Jacobian* and *Hessian* matrices of the implicit surrogate models and finally transfer them into analytical forms. The latter can be easily solved by the existing mature interior point method. Subsequently, SA-TCOP can be efficiently solved to inform operational decision-making.

- Comparison results show that our proposed method can achieve accurate TTC estimation in second and it outperforms the common-used heuristic methods [10]–[12], [21] with around three-orders of magnitude improvement in decision-making speed in the rolling horizon operational planning.

## II. TTC CONSTRAINED OPERATIONAL PLANNING

In this section, the TCOP model is presented, whose objective is to minimize the operation costs while meeting several regular constraints and TTC security constraints. Note that, to deal with uncertainties from wind generations and loads, the TCOP model is formulated in the model predictive control (MPC) form. It will be demonstrated later in Section III that this MPC form naturally provides the interfaces for surrogate model updating. MPC refers to a sequence of optimized operations, where every iteration of the optimization is motivated by the current response determined by the last optimization operation and the predicted information from a specified future horizon [22], [23].

*1) MPC formulation of the models:* Let us first define the optimization variables (OVs). Assume that the operating condition at the time $kT_s$ is available from either power flow or state estimation, where $T_s$ is the optimization time step, the OVs at the time $(k+1)T_s$ can be determined by taking the $k$th operating condition as the initial scenario and considering the future uncertainties over the horizon $[(k+1)T_s, (k+u)T_s]$. These variables include the control variables (i.e., active power outputs of generators) and state variables (i.e., bus voltages and phase angles). Hence, we define the OVs over the horizon $[(k+1)T_s, (k+u)T_s]$ as (1), and $T_s$ is omitted here for simplification:

$$\boldsymbol{x}|_k^{[k+1,k+u]} \triangleq [\boldsymbol{x}_{\text{con}}; \ \boldsymbol{x}_{\text{state}}]|_k^{[k+1,k+u]}, \boldsymbol{x} \in R^{nu \times 1}, k \in \mathbb{T} \quad (1)$$

where $n$ is the dimension of the OVs vector at one period; and $\mathbb{T}$ is the optimization time interval.

*2) Objective function:* it aims to minimize the generation cost as well as the wind curtailment cost over a horizon, i.e.,

$$\text{Minimize} \sum_{t=k+1}^{k+u} \Delta t \left[ \sum_{g \in \mathbb{G}} C_g(P_g(t)) + \sum_{w \in \mathbb{W}} C_w(\Delta P_w(t)) \right] \quad (2a)$$

$$C_g(P_g) = a_g P_g^2 + b_g P_g + c_g \quad (2b)$$

$$C_w(\Delta P_w) = c_w \Delta P_w \quad (2c)$$

where $C_g(P_g)$ is the generation cost function; $a_g$, $b_g$ and $c_g$ are cost coefficients of the $g$-th generator; $C_w(\Delta P_w)$ is the wind curtailment cost function; $c_w$ is the cost per MW of curtailed wind power. $\mathbb{G}$ and $\mathbb{W}$ are the sets of generators and wind farms, respectively.

*3) Generators constraints:*

$$P_g(0) = P_g^*(k) \quad (3a)$$

$$P_g(t) + \Delta P_g(t) = P_g(t+1) \quad (3b)$$

$$\Delta P_g^{\min}(t) < \Delta P_g(t) < \Delta P_g^{\max}(t) \quad (3c)$$

$$P_g^{\min}(t+1) < P_g(t+1) < P_g^{\max}(t+1) \quad (3d)$$

$$Q_g^{\min}(t+1) < Q_g(t+1) < Q_g^{\max}(t+1) \quad (3e)$$

$$\forall t \in \{0, 1, \dots, u-1\}, \forall g \in \mathbb{G}, k \in \mathbb{T}$$

where (3a) updates the initial operating condition of the current MPC horizon to the optimized ones of the last horizon.

*4) Power flow constraints:*

$$P_i^{Inj}(t) - V_i(t) \sum_{j \in i} V_j(t)[G_{ij}\cos(\theta_{ij,t}) + B_{ij}\sin(\theta_{ij,t})] = 0 \quad (4a)$$

$$Q_i^{Inj}(t) - V_i(t) \sum_{j \in i} V_j(t)[G_{ij}\sin(\theta_{ij,t}) - B_{ij}\cos(\theta_{ij,t})] = 0 \quad (4b)$$

$$\forall t \in \{k+1, k+2, \dots k+u\}, \forall i, j \in \mathbb{B}, k \in \mathbb{T}$$

where $\mathbb{B}$ denotes the set of buses.

*5) Operational constraints:*

$$V_i^{\min}(t) < V_i(t) < V_i^{\max}(t), \forall i \in \mathbb{B} \quad (5a)$$

$$P_{ij,l}^{\min}(t) < P_{ij,l}(t) < P_{ij,l}^{\max}(t), \forall i, j \in \mathbb{B} \cap \mathbb{L}, l \in \mathbb{L} \quad (5b)$$

$$\forall t \in \{k+1, k+2, \dots k+u\}, k \in \mathbb{T}$$

where $\mathbb{L}$ denotes the set of lines.

*6) Wind power constraints:*

$$0 \leq \Delta P_i(t) \leq P_i^{\text{Ref}}(t) \quad (6a)$$

$$P_i^{\text{Ref}}(t) = \begin{cases} P_i^{\text{Ref}}(t), & 0 \leq P_i^{\text{Ref}}(t) < P_i^{\max}(t) \\ P_i^{\max}(t), & P_i^{\text{Ref}}(t) \geq P_i^{\max}(t) \end{cases} \quad (6b)$$

$$P_i(t) = P_i^{\text{Ref}}(t) - \Delta P_i(t) \quad (6c)$$

$$P_i^{\min}(t) \leq P_i(t) \leq P_i^{\max}(t) \quad (6d)$$

$$\forall t \in \{k+1, k+2, \dots k+u\}, \forall i \in \mathbb{W}, k \in \mathbb{T}$$

where $P_i^{\text{Ref}}(t)$ is the forecast of the $i$-th wind farm output at period $t$.

*7) Energy storage system (ESS) constraints:*

$$-P_i^{\text{Char,max}}(t+1) \leq P_i(t+1) \leq P_i^{\text{Dischar,max}}(t+1) \quad (7a)$$

$$E_i(0) = E_i^*(k) \quad (7b)$$

$$E_i(t+1) \leq E_i(t) + \Delta t P_i^{\text{Char,max}}(t+1) \quad (7c)$$

$$E_i^{\min}(t+1) \leq E_i(t+1) \leq E_i^{\max}(t+1) \quad (7d)$$

$$\forall t \in \{0, 1, \dots, u-1\}, \forall i \in \mathbb{E}, k \in \mathbb{T}$$

where $\mathbb{E}$ is the set of ESSs.

*8) TTC security constraints:* For each tie-line, the power flow should be controlled to be below the TTC limit with a user-defined target margin $\tau$ ($\tau \geq 0$):

$$\eta_l(t) \triangleq P_{ij,l}(t) - \Gamma_{ij,l}(t) \leq -\epsilon \quad (8)$$

$$\forall t \in \{k+1, k+2, \dots k+u\}, \forall l \in \mathbb{K}, i, j \in \mathbb{B} \cap \mathbb{K}$$

where $\mathbb{K}$ denotes a set of tie-lines; $\eta_l(t)$ is the security margin of the $l$-th tie-line at time $t$; $\Gamma_{ij,l}(t)$ denotes the transfer capability of the $l$-th tie-line linked with buses $i$ and $j$ at time $t$. For TTC security constraints, $\epsilon$ is conservatively set to be 0.05. $\Gamma_{ij,l}(t)$ is a dynamic variable that needs to be determined by solving DSCOPF. Several model-based methods have been proposed to calculate TTC [3], [16], [17], [24], but the repeated power flow (RPF)-based method is highly compatible with sophisticated DSCOPF [16], [17] and therefore used in this paper in the training stage.

The RPF-based TTC calculation model is formulated below.

$$\boldsymbol{\mathcal{X}}(\tau_0) = \boldsymbol{x}_{\text{con}}(t), \boldsymbol{\mathcal{Y}}(\tau_0) = \boldsymbol{x}_{\text{state}}(t) \quad (9)$$

$$\text{Maximize } \lambda_t \quad (10a)$$

$$\text{s.t. } \boldsymbol{H}_c(\boldsymbol{\mathcal{X}}(\tau_0), \boldsymbol{\mathcal{Y}}(\tau_0), \lambda_t) = \boldsymbol{0}, \forall c \in c_0 \cup \mathcal{C} \quad (10b)$$

$$\boldsymbol{V}_c(\boldsymbol{\mathcal{X}}(\tau_0), \boldsymbol{\mathcal{Y}}(\tau_0), \lambda_t) \leq \boldsymbol{0}, \forall c \in c_0 \cup \mathcal{C} \quad (10c)$$

$$\begin{cases} \dot{\boldsymbol{\mathcal{X}}}(\tau) = \boldsymbol{\mathcal{G}}_c(\boldsymbol{\mathcal{X}}(\tau), \boldsymbol{\mathcal{Y}}(\tau), \lambda_t), \forall c \in \boldsymbol{\mathcal{C}}, \tau \in [\tau_0, \tau_{\text{end}}] \\ \psi_c(\boldsymbol{\mathcal{X}}(\tau), \boldsymbol{\mathcal{Y}}(\tau), \lambda_t) \le \boldsymbol{0}, \forall c \in \boldsymbol{\mathcal{C}}, \tau \in [\tau_0, \tau_{\text{end}}] \end{cases} \quad (10\text{d})$$

where $(\boldsymbol{\mathcal{X}}, \boldsymbol{\mathcal{Y}})$ are the variables in the model (10); (9) and (10) imply that TTC at period $t$ will be calculated based on the current operating condition; $\lambda_t$ is utilized to increase the load of sink area or the generation of source area [17]; $\boldsymbol{\mathcal{C}}$ denotes the set of contingencies, while $c_0$ implies the operation condition without contingency; $[\boldsymbol{\mathcal{X}}(\tau), \boldsymbol{\mathcal{Y}}(\tau)]$ denotes the operating condition during the transient response period $(\tau_0, \tau_{\text{end}})$; (10d) represents dynamic security constraints. In this paper, the transient stability constraints are considered and $\psi_c(\cdot)$ is the transient stability criterion adopted in this paper. It is shown as follows [3]:

$$\begin{cases} |\delta_i(\tau) - \delta_{\text{COI}}(\tau)| - \delta_{\max} < 0 \\ \delta_{\text{COI}}(\tau) = \sum_i \mathrm{M}_i \, \delta_i(\tau) / \sum_i \mathrm{M}_{i'} \end{cases}, \forall i \in \mathbb{G}, \tau \in [\tau_0, \tau_{\text{end}}] \quad (11)$$

By using algorithm 1 in [17], TTC can be obtained via

$$\begin{cases} \lambda_t^* = \text{argmax } \lambda_t \\ \Gamma_{lj,l}(t) = PF_l([\boldsymbol{\mathcal{X}}(\tau_0), \boldsymbol{\mathcal{Y}}(\tau_0)], \lambda_t^*), \forall l \in \mathbb{K} \end{cases} \quad (12)$$

where $PF_l(\cdot)$ is the function for calculating the power transfer of line $l$. In the process of simulating (10d), the wind turbines and generators are modeled by DFIGs and 4-th order models in PSAT [25], respectively. Note that since this paper focuses on preventive control strategies considering TTC security constraints for economic dispatch, ESSs are modeled by a simple first-order dynamic model as shown in [26].

## III. Proposed Surrogate-assisted Approach

The TCOP formed in the previous section is indeed a complicated and computationally intensive problem, which is challenging for existing methods. Here, a surrogate model-based method is proposed to deal with that.

### A. Offline Construction of TTC Estimator/Surrogate Model

To build the data-driven TTC surrogate model, substantial stochastic scenarios are fully sampled, in which TTC is computed via the model-based RPF model. Nonlinear learning algorithms are then utilized to learn the stochastic dynamic patterns from these samples in a supervised manner. The detailed steps are as follows:

First of all, we define the input features $\boldsymbol{\mathcal{U}}^S$ and the target features $\boldsymbol{Y}^S$ by (13):

$$\boldsymbol{\mathcal{U}}^S = \{\boldsymbol{P}_i^{Inj}, \boldsymbol{Q}_i^{Inj}, \boldsymbol{V}_g, \boldsymbol{P}_j^{Load}, \boldsymbol{Q}_j^{Load}, \boldsymbol{V}_b\}, \quad (13\text{a})$$

$$\boldsymbol{Y}^S = \{\Gamma_l\}, \forall i \in \mathbb{G} \cup \mathbb{W} \cup \mathbb{E}, g \in \mathbb{G}, j \in \mathbb{B}, l \in \mathbb{K} \quad (13\text{b})$$

Define a data collector $\boldsymbol{\mathcal{D}}(\cdot)$ to collect $\boldsymbol{\mathcal{U}}^S$ from a certain operation condition, i.e., $\boldsymbol{\mathcal{U}}^S = \boldsymbol{\mathcal{D}}(\boldsymbol{\mathcal{X}}, \boldsymbol{\mathcal{Y}})$, where $(\boldsymbol{\mathcal{X}}, \boldsymbol{\mathcal{Y}})$ is used to denote the initial operation condition of TTC calculation. Then, samples are generated via (14):

$$\begin{cases} \boldsymbol{v}_{con}^s = \boldsymbol{v}_{con}^{\min} + \{r_{j1} \cdot n\} \times (\boldsymbol{v}_{con}^{\max} - \boldsymbol{v}_{con}^{\min}) \\ \boldsymbol{v}_{con} = \{\boldsymbol{P}_i^{Inj}, \boldsymbol{Q}_i^{Inj}, \boldsymbol{V}_g\}, \forall i \in \mathbb{G} \cup \mathbb{W} \cup \mathbb{E}, g \in \mathbb{G} \end{cases} \quad (14\text{a})$$

$$\begin{cases} \boldsymbol{v}_{Load}^s = \boldsymbol{\mu}^{load} - 2\varrho\boldsymbol{\sigma}^{Load} + \{r_{j2} \cdot n\} \times 2\varrho\boldsymbol{\sigma}^{Load} \\ \boldsymbol{v}_{Load} = \{\boldsymbol{P}_i^{Load}, \boldsymbol{Q}_i^{Load}\}, \forall i \in \mathbb{B} \end{cases} \quad (14\text{b})$$

$$j1 = 1, \dots, \#_{con}; j2 = 1, \dots, \#_{Load}; n = 1, 2, \dots, n^{IN}$$

where $r_{j1}$ and $r_{j2}$ equal to $e^{j1}$ and $e^{j2}$, respectively; $\#_{con}$ and $\#_{Load}$ are respectively the dimensions of $\boldsymbol{v}_{con}$ and $\boldsymbol{v}_{Load}$; $\boldsymbol{v}_{con}^{\max}$ and $\boldsymbol{v}_{con}^{\min}$ are respectively the restricted upper and lower bounds of $\boldsymbol{v}_{con}$; $\boldsymbol{\mu}^{Load}$ and $\boldsymbol{\sigma}^{Load}$ are the mean values and maximum deviations of historical loads, and $\varrho$ ($\varrho > 1$) is a coefficient used to expand the sampling space of loads to ensure the coverage of actual operating space. For each sample $[\boldsymbol{v}_{con}^s, \boldsymbol{v}_{Load}^s]$, we perform power flow program to get $(\boldsymbol{\mathcal{X}}^s, \boldsymbol{\mathcal{Y}}^s)$, then $\boldsymbol{\mathcal{U}}^S = \boldsymbol{\mathcal{D}}(\boldsymbol{\mathcal{X}}^s, \boldsymbol{\mathcal{Y}}^s)$.

To attain $\boldsymbol{Y}^S$, (12) is used by passing $(\boldsymbol{\mathcal{X}}^S, \boldsymbol{\mathcal{Y}}^S)$ through it. Afterward, the sample set $[\boldsymbol{\mathcal{U}}^S, \boldsymbol{Y}^S]$ can be then produced. They are further used by nonlinear learning algorithms to obtain TTC estimator/surrogate model. Here, we briefly introduce 3 algorithms, i.e., elastic net (EN), single hidden layer NN (SLNN), and DLNN.

#### 1) Elastic Nets

EN usually exhibits salient performance in different applications among manifold linear regression techniques [27]. Considering that it has been introduced to some power system applications [15], EN is taken as a comparative method in this paper.

EN employs both $L_1$ and $L_2$ norms. For building a TTC estimator for the $l$-th tie-line, the $Loss$ function is formed as (15):

$$Loss_l(\boldsymbol{\beta}_l) = \|\boldsymbol{Y}^S - \boldsymbol{\mathcal{U}}^S\boldsymbol{\beta}_l\|_2^2 + \gamma_1^{EN}\|\boldsymbol{\beta}_l\|_1 + \gamma_2^{EN}\|\boldsymbol{\beta}_l\|_2^2, l \in \mathbb{K} \quad (15)$$

By solving (15), the EN based TTC estimator can be obtained as shown in (16):

$$\begin{cases} \boldsymbol{\beta}_l^* = \text{arg } \min_{\boldsymbol{\beta}_l} Loss_l(\boldsymbol{\beta}_l) \\ \tilde{\Gamma}_l(t) = \Phi_l^{EN}[\boldsymbol{\mathcal{D}}(\boldsymbol{x}(t))] \triangleq \boldsymbol{\mathcal{D}}(\boldsymbol{x}(t))\boldsymbol{\beta}_l^* \end{cases}, \forall l \in \mathbb{K} \quad (16)$$

where $\tilde{\Gamma}_l(t)$ indicates the TTC estimation via surrogate model $\Phi_l^{EN}$. Upon the definition of $\boldsymbol{\mathcal{D}}(\cdot)$, $\boldsymbol{\mathcal{D}}(\boldsymbol{x}(t))$ denotes that the input to a surrogate is collected from the optimization variables at $t$.

#### 2) Single Hidden Layer Neural Networks

The SLNNs are constructed by fully connecting the input layer, hidden layer, and output layer. To represent the structure of SLNNs, we denote:

$$\boldsymbol{\mathcal{M}}_\ell(\boldsymbol{x}_\ell) = \boldsymbol{w}_\ell \boldsymbol{x}_\ell + \boldsymbol{b}_\ell, \boldsymbol{w}_\ell \in \mathbb{R}^{\mathcal{N}_\ell \times \mathcal{N}_{\ell-1}} \quad (17)$$

$$\begin{cases} \boldsymbol{w}_\ell = [\boldsymbol{w}_\ell^1; \dots; \boldsymbol{w}_\ell^m; \dots; \boldsymbol{w}_\ell^{\mathcal{N}_\ell}], \boldsymbol{w}_\ell^m \in \mathbb{R}^{1 \times \mathcal{N}_{\ell-1}} \\ \boldsymbol{b}_\ell = [b_\ell^1; \dots; b_\ell^m; \dots; b_\ell^{\mathcal{N}_\ell}] \end{cases} \quad (18)$$

where $\mathcal{N}_i$ denotes the dimension of the input for the $i$-th layer; $i = 0$ represents the dimension of the input features. $m$ denotes the weights connecting $m$-th hidden neurons and the output of the previous layer. In this paper, the activation function $\mathcal{S}(x)$ of the hidden layer can be either the *Sigmoid* function or the *Softplus* function. They are represented as follows:

$$\mathcal{S}(x) = \begin{cases} \dfrac{1}{1 + e^{-2x}} - 1 & , if \ Sigmoid \\ \ln(1 + e^x) & , if \ Softplus \end{cases} \quad (19)$$

The output of the SLNNs can be computed via (20):

$$\Phi^{SL}(\boldsymbol{\mathcal{U}}^s) \triangleq \mathcal{O}\left(\mathcal{S}\left(\mathcal{M}_1(\boldsymbol{\mathcal{U}}^s)\right)\right) \quad (20)$$

where $\mathcal{O}$ is the output function, which is a linear sum function defined as: $\mathcal{O}(\boldsymbol{x}) = \mathcal{M}_{\mathcal{L}}(\boldsymbol{x}_{\mathcal{L}})$, where $\mathcal{L}$ is the number of all layers (including output layer) of the NN. $\mathcal{L}=2$ is used for an SLNN.

The $L_2$ norm is included in the $Loss$ function, and the Broyden, Fletcher, Goldfarb, and Shanno (BFGS) quasi-newton algorithm [28], is employed to optimize the SLNNs by (21):

$$\underset{\boldsymbol{w}_\ell, \boldsymbol{b}_\ell \text{ [BFGS]}}{\text{Minimize}} Loss = \gamma^{SL}\|\boldsymbol{Y}^S - \Phi^{SL}(\boldsymbol{\mathcal{U}}^s)\|_2^2 + (1 - \gamma^{SL})\sum_{\ell=1}^{\mathcal{L}}\|\boldsymbol{w}_\ell\|_2^2 \quad (21)$$

Then, TTC surrogate model is represented by (22):

$$\tilde{\Gamma}_l(t) = \Phi_l^{SL}[\boldsymbol{\mathcal{D}}(\boldsymbol{x}(t))], \forall l \in \mathbb{K} \quad (22)$$

#### 3) Deep-Hidden Layer Neural Networks

A $\{\mathcal{N}_{\mathcal{L}} - \mathcal{N}_{\mathcal{L}-1} - \cdots \mathcal{N}_1\}$ DLNN can be constructed by (23):

$$\Phi^{\mathcal{L}DL}(\boldsymbol{\mathcal{U}}^s) \triangleq \mathcal{O}\left(\mathcal{S}\left(\mathcal{M}_{\mathcal{L}-1}(\dots \mathcal{S}(\mathcal{M}_1(\boldsymbol{\mathcal{U}}^s))\dots)\right)\right) \quad (23)$$

The $Loss$ function of such DLNN is the weighted sum of the estimated error and $L_2$ norm. In this paper, the deep belief networks (DBNs) are leveraged to train the DLNN [29], [30]. It is shown in (24):

$$\underset{\boldsymbol{w}_\ell, \boldsymbol{b}_\ell \text{ [DBN]}}{\text{Minimize}} \gamma^{\text{DL}} \|\boldsymbol{Y}^s - \Phi^{\mathcal{L}\text{DL}}(\boldsymbol{\mathcal{U}}^s)\|_2^2 + (1 - \gamma^{\text{DL}}) \sum_{\ell=1}^{\mathcal{L}} \|\boldsymbol{w}_\ell\|_2^2 \quad (24)$$

After that, the TTC surrogate model is represented by (25), i.e.,

$$\tilde{t}_l(t) = \Phi_l^{\mathcal{L}\text{DL}}\big[\mathcal{D}(\boldsymbol{x}(t))\big], \forall l \in \mathbb{K} \quad (25)$$

It should be noted that the sample generation procedure and the training process of the surrogate models are both offline. Then, the properly built surrogates can be used for online TTC calculation.

### B. Surrogate Model Assisted TCOP Reformulation

To reduce the complexity incurred by solving (8)~(12) in the original TCOP model, the constructed TTC surrogate model is used to replace (8)~(12). Accordingly, the reformulated SA-TCOP model can be expressed as follows:

$$\underset{\boldsymbol{x}}{\text{Minimize}} \ (2) \quad (26a)$$

$$s.t. \ (3) \sim (7) \quad (26b)$$

$$\tilde{\eta}_l(t) \triangleq P_{ij,l}(t) - \Phi_l^{\mathcal{P}}\big[\mathcal{D}(\boldsymbol{x}(t))\big] \leq -\tau \quad (26c)$$

$$\forall t \in \{k+1, k+2, \dots k+u\}, \forall l \in \mathbb{K}, i, j \in \mathbb{B} \cap \mathbb{K}$$

where the superscript $\mathcal{P}$ denotes the type of surrogate models. With the surrogate model, the model-based constraints (8) are converted into the computationally tractable constraints (26c). However, (26) is still of high-dimensionality and it is necessary to devise a gradient algorithm to deal with (26). This is shown in the next section.

## IV. PROPOSED SOLUTION METHOD

In this section, the *Jacobian* and *Hessian* matrices of the adopted surrogate models versus the input will be deduced. This allows us to deal with (26c) analytically.

### A. Interior Point Method

The SA-TCOP model shown in (26) can be organized into the following compact form:

$$\text{Minimize} \ F(\boldsymbol{x}_{\text{con}}) \quad (27a)$$

$$s.t. \ \boldsymbol{H}(\boldsymbol{x}) = \boldsymbol{0} \quad (27b)$$

$$[\boldsymbol{g}(\boldsymbol{x}) \leq \boldsymbol{0}] \triangleq \begin{cases} \boldsymbol{g}_p(\boldsymbol{x}) \leq \boldsymbol{0} \\ \boldsymbol{g}_s(\boldsymbol{x}) \leq \boldsymbol{0} \end{cases} \quad (27c)$$

where $\boldsymbol{H}(\boldsymbol{x})$ denotes the equality constraints; $\boldsymbol{g}_p(\boldsymbol{x})$ and $\boldsymbol{g}_s(\boldsymbol{x})$ represent the constraints for the original physical models in (3)-(7) and the SA in (26c), respectively.

The process of solving (27) by IPM is as follows: *1)* the positive slack vectors $\boldsymbol{l}^s$ and $\boldsymbol{u}^s$ are introduced to turn $\boldsymbol{g}(\boldsymbol{x})$ into equality constraints; *2)* the barrier parameter $\mu$ ($\mu > 0$) is used to transfer $F(\boldsymbol{x})$ into a barrier function; and *3)* with Lagrange multipliers $\boldsymbol{y}, \boldsymbol{z}^L$ ($\boldsymbol{z}^L > 0$), and $\boldsymbol{w}^L$ ($\boldsymbol{w}^L > 0$), the Lagrange function can be built as (28):

$$L = F(\boldsymbol{x}) - (\boldsymbol{z}^L)^{\text{T}} \big[\boldsymbol{g}(\boldsymbol{x}) - \boldsymbol{l}^s - \underline{\boldsymbol{g}}\big] - (\boldsymbol{w}^L)^{\text{T}} [\boldsymbol{g}(\boldsymbol{x}) + \boldsymbol{u}^s - \overline{\boldsymbol{g}}]$$
$$- \boldsymbol{y}^{\text{T}} \boldsymbol{H}(\boldsymbol{x}) - \mu \left[ \sum_{i=1}^{r} \log(l_i^s) + \sum_{i=1}^{r} \log(u_i^s) \right] \quad (28)$$

where $r$ is the number of inequality constraints.

*4)* Considering the perturbed KKT conditions and utilizing the Newton-Raphson method, the correction equation is determined by (29):

$$\begin{bmatrix} \mathcal{H} & \nabla \boldsymbol{H}(\boldsymbol{x}) \\ \nabla \boldsymbol{H}(\boldsymbol{x})^{\text{T}} & \boldsymbol{0} \end{bmatrix} \begin{bmatrix} \Delta \boldsymbol{x} \\ \Delta \boldsymbol{y} \end{bmatrix} = - \begin{bmatrix} \boldsymbol{\psi} \\ \boldsymbol{H} \end{bmatrix} \quad (29)$$

where $\mathcal{H}$ is shown in (30), and $\boldsymbol{\psi}$ can be found in [2].

$$\mathcal{H} = \boldsymbol{y}^{\text{T}} \nabla^2 \boldsymbol{H}(\boldsymbol{x}) + [(\boldsymbol{z}^L)^{\text{T}} + (\boldsymbol{w}^L)^{\text{T}}] \nabla^2 \boldsymbol{g}(\boldsymbol{x}) - \nabla^2 F(\boldsymbol{x}) + \nabla \boldsymbol{g}(\boldsymbol{x})(\boldsymbol{w}^L \oslash \boldsymbol{u}^s - \boldsymbol{z}^L \oslash \boldsymbol{l}^s)[\nabla \boldsymbol{g}(\boldsymbol{x})]^{\text{T}} \quad (30)$$

Furthermore, we can calculate $\Delta \boldsymbol{l}^s, \Delta \boldsymbol{u}^s, \Delta \boldsymbol{w}^L$, and $\Delta \boldsymbol{z}^L$ by (31):

$$\begin{cases} \Delta \boldsymbol{z}^L = -(\boldsymbol{L}^s)^{-1} \boldsymbol{L}_{is}^\mu - (\boldsymbol{L}^s)^{-1} \boldsymbol{Z}^L \Delta \boldsymbol{l}^s \\ \Delta \boldsymbol{l}^s = \boldsymbol{L}_z + \nabla_{\boldsymbol{x}}^{\text{T}} \boldsymbol{g}(\boldsymbol{x}) \Delta \boldsymbol{x} \\ \Delta \boldsymbol{w}^L = -(\boldsymbol{U}^s)^{-1} \boldsymbol{L}_{\boldsymbol{u}^s}^\mu - (\boldsymbol{U}^s)^{-1} \boldsymbol{W}^L \Delta \boldsymbol{u}^s \\ \Delta \boldsymbol{u}^s = -\boldsymbol{L}_w - \nabla_{\boldsymbol{x}}^{\text{T}} \boldsymbol{g}(\boldsymbol{x}) \Delta \boldsymbol{x} \end{cases} \quad (31)$$

where $\boldsymbol{L}^s, \boldsymbol{U}^s, \boldsymbol{Z}^L$ and $\boldsymbol{W}^L$ are diagonal matrices with elements of $\boldsymbol{l}^s, \boldsymbol{u}^s, \boldsymbol{z}^L$ and $\boldsymbol{w}^L$, respectively.

*5)* Update the primal and dual variables by (32) and (33):

$$\alpha_{\text{p}} = 0.9995 \left[ \min \left( \frac{-l_i^s}{\Delta l_i^s}, \Delta l_i^s < 0; \frac{-u_i^s}{\Delta u_i^s}, \Delta u_i^s < 0 \right), 1 \right] \quad (32a)$$

$$\alpha_{\text{d}} = 0.9995 \left[ \min \left( \frac{-z_i}{\Delta z_i}, \Delta z_i < 0; \frac{-w_i}{\Delta w_i}, \Delta w_i > 0 \right), 1 \right] \quad (32b)$$

$$\begin{cases} \boldsymbol{x}^{(k+1)} = \boldsymbol{x}^{(k)} + \alpha_{\text{p}} \Delta \boldsymbol{x}, \boldsymbol{y}^{(k+1)} = \boldsymbol{y}^{(k)} + \alpha_{\text{d}} \Delta \boldsymbol{y} \\ \boldsymbol{l}^{s,(k+1)} = \boldsymbol{l}^{s,(k)} + \alpha_{\text{p}} \Delta \boldsymbol{l}^s, \boldsymbol{u}^{s,(k+1)} = \boldsymbol{u}^{s,(k)} + \alpha_{\text{p}} \Delta \boldsymbol{u}^s \\ \boldsymbol{z}^{L,(k+1)} = \boldsymbol{z}^{L,(k)} + \alpha_{\text{d}} \Delta \boldsymbol{z}^L, \boldsymbol{w}^{L,(k+1)} = \boldsymbol{w}^{L,(k)} + \alpha_{\text{d}} \Delta \boldsymbol{w}^L \end{cases} \quad (33)$$

*6)* Update $\mu$ by (34):

$$\mu = \sigma \frac{Gap}{2r} = \sigma \frac{(\boldsymbol{l}^s)^{\text{T}} \boldsymbol{z}^L - (\boldsymbol{u}^s)^{\text{T}} \boldsymbol{w}^L}{2r} \quad (34)$$

where $\sigma$ is the central parameter.

*7)* Terminating criterion: if $Gap < \varepsilon$, then output the current $\boldsymbol{x}$; otherwise execute *4)-6)*. Here $\varepsilon$ is a specified positive threshold.

### B. Deriving Analytic Surrogate Model for IPM Method

IPM needs the gradient information of $\boldsymbol{g}$. $\nabla \boldsymbol{g}_p$ and $\nabla^2 \boldsymbol{g}_p$ can be easily obtained but this is not the case for $\nabla \boldsymbol{g}_s$ and $\nabla^2 \boldsymbol{g}_s$. This is because the surrogate model is in the "black-box" form. From (26c), $\nabla \tilde{\eta}_l(t)$ and $\nabla^2 \tilde{\eta}_l(t)$ can be calculated by (35):

$$\begin{cases} \nabla \tilde{\eta}_l(t) = \nabla P_{ij,l}(t) - \nabla \Phi_l^{\mathcal{P}}\big[\mathcal{D}(\boldsymbol{x}(t))\big] \\ \nabla^2 \tilde{\eta}_l(t) = \nabla^2 P_{ij,l}(t) - \nabla^2 \Phi_l^{\mathcal{P}}\big[\mathcal{D}(\boldsymbol{x}(t))\big] \end{cases} \quad (35)$$

(35) indicates that $\nabla \boldsymbol{g}_s$ and $\nabla^2 \boldsymbol{g}_s$ can be readily derived as long as $\nabla \Phi_l^{\mathcal{P}}$ and $\nabla^2 \Phi_l^{\mathcal{P}}$ can be obtained. Note that $\nabla \Phi_l^{\mathcal{P}}$ and $\nabla^2 \Phi_l^{\mathcal{P}}$ are different for different surrogate models. We derive the related forms for the aforementioned three surrogate models below.

*1) Elastic Nets*

From (16), $\nabla \Phi_l^{\text{EN}}$ and $\nabla^2 \Phi_l^{\text{EN}}$ can be deduced as:

$$\begin{cases} \nabla_{\boldsymbol{x}} \Phi_l^{\text{EN}}(\mathcal{D}(\boldsymbol{x})) = \frac{\partial \mathcal{D}(\boldsymbol{x})}{\partial \boldsymbol{x}} \boldsymbol{\beta}_l^* , \forall l \in \mathbb{K}, \mathcal{D}(\boldsymbol{x}) \in \text{R}^{N_0 \times 1} \\ \nabla_{\boldsymbol{x}}^2 \Phi_l^{\text{EN}}(\mathcal{D}(\boldsymbol{x})) = \boldsymbol{0} \end{cases} \quad (36)$$

*2) Single Hidden-Layer Neural Networks*

The derivative of $\mathcal{S}(x)$ versus $x$ is given in (37):

$$\begin{cases} \text{d}_x \mathcal{S}(x) \equiv \frac{\text{d}\mathcal{S}(x)}{\text{d}x} = \frac{4e^{-2x}}{(1+e^{-2x})^2} \\ \text{d}_{x,x}^2 \mathcal{S}(x) \equiv \frac{\text{d}^2 \mathcal{S}(x)}{\text{d}x^2} = -2\mathcal{S}(x)\text{d}_x \mathcal{S}(x) \end{cases}, if \ Sigmoid \quad (37a)$$

$$\begin{cases} \text{d}_x \mathcal{S}(x) \equiv \frac{\text{d}\mathcal{S}(x)}{\text{d}x} = \frac{e^x}{1+e^x} \\ \text{d}_{x,x}^2 \mathcal{S}(x) \equiv \frac{\text{d}^2 \mathcal{S}(x)}{\text{d}x^2} = \text{d}_x \mathcal{S}(x)(1-\text{d}_x \mathcal{S}(x)) \end{cases}, if \ Softplus \quad (37b)$$

For simplification, $\boldsymbol{w}_\ell$ and $\boldsymbol{b}_\ell$ respectively represent the weight and bias of properly trained NN, and $\mathcal{D}$ implies $\mathcal{D}(\boldsymbol{x}(t))$. Then, based on the chain rule, the *Jacobian* matrix is

$$\frac{\partial \bar{l}_l(t)}{\partial x_i(t)} = \frac{d\bar{l}_l(t)}{d\mathcal{S}(\mathcal{M}_1(\boldsymbol{\mathcal{D}}))} \times \frac{d\mathcal{S}(\mathcal{M}_1(\boldsymbol{\mathcal{D}}))}{d\mathcal{M}_1(\boldsymbol{\mathcal{D}})} \times \frac{d\mathcal{M}_1(\boldsymbol{\mathcal{D}})}{d\boldsymbol{\mathcal{D}}} \times \frac{\partial \boldsymbol{\mathcal{D}}}{\partial x_i(t)}$$
$$= \boldsymbol{w}_2[d_x \mathcal{S}(\mathcal{M}_1(\boldsymbol{\mathcal{D}})) \times (\boldsymbol{w}_1 \boldsymbol{e}_i)] \quad (38)$$
$$\boldsymbol{w}_1 \in \mathbb{R}^{\mathcal{N}_1 \times \mathcal{N}_0}, \boldsymbol{w}_2 \in \mathbb{R}^{1 \times \mathcal{N}_1}, \boldsymbol{e}_i \in \mathbb{R}^{\mathcal{N}_0 \times 1}$$

where $x_i(t)$ is the i-th variable at period $t$; $\boldsymbol{e}_i$ is an identity vector in which only the $i$-th element is 1.

The formulation of $d_x \mathcal{S}(\mathcal{M}_1(\boldsymbol{\mathcal{D}}))$ is shown as (39):

$$d_x \mathcal{S}(\mathcal{M}_1(\boldsymbol{\mathcal{D}})) =$$
$$\begin{bmatrix} d_{\mathcal{M}_1^1(\boldsymbol{\mathcal{D}})} \mathcal{S}(\mathcal{M}_1^1(\boldsymbol{\mathcal{D}})) & 0 & 0 \\ 0 & \ddots & 0 \\ 0 & 0 & d_{\mathcal{M}_1^{\mathcal{N}_1}(\boldsymbol{\mathcal{D}})} \mathcal{S}(\mathcal{M}_1^{\mathcal{N}_1}(\boldsymbol{\mathcal{D}})) \end{bmatrix} \quad (39a)$$

$$\begin{cases} \mathcal{M}_1(\boldsymbol{\mathcal{D}}) = [\mathcal{M}_1^1(\boldsymbol{\mathcal{D}}); \dots; \mathcal{M}_1^{\mathcal{N}_1}(\boldsymbol{\mathcal{D}})] \in \mathbb{R}^{\mathcal{N}_1 \times 1} \\ \mathcal{M}_1^m(\boldsymbol{\mathcal{D}}) = \boldsymbol{w}_1^m \boldsymbol{\mathcal{D}} + b_1^m \end{cases} \quad (39b)$$

Then, based on the above derivatives, we can get the *Hessian* matrices of SLNN with respect to the inputs as

$$\frac{\partial^2 \bar{l}_l(t)}{\partial x_i(t) \partial x_j(t)}$$
$$= \frac{d\bar{l}_l(t)}{d\mathcal{S}(\mathcal{M}_1(\boldsymbol{\mathcal{D}}))} \times \frac{d^2\mathcal{S}(\mathcal{M}_1(\boldsymbol{\mathcal{D}}))}{d\mathcal{M}_1(\boldsymbol{\mathcal{D}}) \partial x_j(t)} \times \frac{d\mathcal{M}_1(\boldsymbol{\mathcal{D}})}{d\boldsymbol{\mathcal{D}}} \times \frac{\partial \boldsymbol{\mathcal{D}}}{\partial x_i(t)}$$
$$= \boldsymbol{w}_2[d^2_{\mathcal{M}_1(\boldsymbol{\mathcal{D}}), x_j} \mathcal{S}(\mathcal{M}_1(\boldsymbol{\mathcal{D}})) \times (\boldsymbol{w}_1 \boldsymbol{e}_i)] \quad (40)$$

In (40), $d^2_{\mathcal{M}_1(\boldsymbol{\mathcal{D}}), x_j} \mathcal{S}(\mathcal{M}_1(\boldsymbol{\mathcal{D}}))$ is deduced as (41):

$$d^2_{\mathcal{M}_1(\boldsymbol{\mathcal{D}}), x_j} \mathcal{S}(\mathcal{M}_1(\boldsymbol{\mathcal{D}})) =$$
$$\begin{bmatrix} d^2_{\mathcal{M}_1^1(\boldsymbol{\mathcal{D}}), x_j} \mathcal{S}(\mathcal{M}_1^1(\boldsymbol{\mathcal{D}})) & 0 & 0 \\ 0 & \ddots & 0 \\ 0 & 0 & d^2_{\mathcal{M}_1^{\mathcal{N}_1}(\boldsymbol{\mathcal{D}}), x_j} \mathcal{S}(\mathcal{M}_1^{\mathcal{N}_1}(\boldsymbol{\mathcal{D}})) \end{bmatrix} \quad (41a)$$

$$d^2_{\mathcal{M}_1^m(\boldsymbol{\mathcal{D}}), x_j} \mathcal{S}(\mathcal{M}_1^m(\boldsymbol{\mathcal{D}})) = \frac{d^2 \mathcal{S}(\mathcal{M}_1^m(\boldsymbol{\mathcal{D}}))}{d\mathcal{M}_1^m(\boldsymbol{\mathcal{D}})^2} \times (\boldsymbol{w}_1^m \boldsymbol{e}_j) \quad (41b)$$

*3) Deep-Hidden Layer Neural Networks*

For simplicity, let $\boldsymbol{w}_1^i$ denote $\boldsymbol{w}_1 \boldsymbol{e}_i$ in the following descriptions. The derivations for *Jacobian* and *Hessian* matrices of DLNNs are similar to that of SLNNs. Thus, the formulas of $\nabla \Phi_l^{\mathcal{L}DL}[\boldsymbol{\mathcal{D}}]$ and $\nabla^2 \Phi_l^{\mathcal{L}DL}[\boldsymbol{\mathcal{D}}]$ are directly given.

Firstly, we define two functions to simplify the formulations of $\nabla \Phi_l^{\mathcal{L}DL}[\boldsymbol{\mathcal{D}}]$ and $\nabla^2 \Phi_l^{\mathcal{L}DL}[\boldsymbol{\mathcal{D}}]$:

$$\mathcal{Q}_\ell(\boldsymbol{\mathcal{D}}) = d_x \mathcal{S}(\mathcal{M}_\ell \dots (\boldsymbol{\mathcal{D}}) \dots) \times \boldsymbol{w}_\ell \quad (42)$$

$$\prod_{i=\ell_1}^{\ell_2} \mathcal{Q}_i(\boldsymbol{\mathcal{D}}) = \begin{cases} \mathcal{Q}_{\ell_1} \times \dots \times \mathcal{Q}_{\ell_2}, & \text{if } L > \ell_1 > \ell_2 \\ \mathcal{Q}_{\ell_2} \times \dots \times \mathcal{Q}_{\ell_1}, & \text{if } L > \ell_2 > \ell_1 \\ 1, & \text{if } \ell_1 \geq L \geq \ell_2 \\ \mathcal{Q}_{\ell_1}, & \text{if } \ell_1 = \ell_2 \end{cases} \quad (43)$$

Note that $d_x \mathcal{S}(\mathcal{M}_\ell \dots (\boldsymbol{\mathcal{D}}) \dots)$ can be similarly obtained based on the deduction of (39a). Based on (42) and (43), the *Jacobian* matrix is written as:

$$\frac{\partial \bar{l}_l(t)}{\partial x_i(t)} = \boldsymbol{w}_L[\prod_{i=\mathcal{L}-1}^2 \mathcal{Q}_i(\boldsymbol{\mathcal{D}})] \times [d_x \mathcal{S}(\mathcal{M}_1(\boldsymbol{\mathcal{D}})) \times \boldsymbol{w}_1^i] \quad (44)$$

Using (42)-(44) and the chain rule, the *Hessian* matrix can be obtained as follows:

$$d_{x_j} \mathcal{Q}_\ell(\boldsymbol{\mathcal{D}}) = d^2_{\hat{x}, x_j} \mathcal{S}(\mathcal{M}_\ell \dots (\boldsymbol{\mathcal{D}}) \dots) \times \boldsymbol{w}_\ell$$
$$= \begin{bmatrix} d^2_{\hat{x}, x_j} \mathcal{S}(\mathcal{M}_\ell^1 \dots (\boldsymbol{\mathcal{D}}) \dots) & 0 & 0 \\ 0 & \ddots & 0 \\ 0 & 0 & d^2_{\hat{x}, x_j} \mathcal{S}(\mathcal{M}_\ell^{\mathcal{N}_\ell} \dots (\boldsymbol{\mathcal{D}}) \dots) \end{bmatrix} \times \boldsymbol{w}_\ell \quad (45a)$$

$$d^2_{\hat{x}, x_j} \mathcal{S}(\mathcal{M}_\ell^m \dots (\boldsymbol{\mathcal{D}}) \dots) = \frac{d^2 \mathcal{S}(\mathcal{M}_\ell^m \dots (\boldsymbol{\mathcal{D}}) \dots)}{d(\mathcal{M}_\ell^m \dots (\boldsymbol{\mathcal{D}}) \dots)^2} \times \boldsymbol{w}_\ell^m \times$$
$$\prod_{i=\ell-1}^2 \mathcal{Q}_i(\boldsymbol{\mathcal{D}}) \times [d_x \mathcal{S}(\mathcal{M}_1(\boldsymbol{\mathcal{D}})) \times \boldsymbol{w}_1^j] \quad (45b)$$

where, in (45), if $\ell = 1$, $\boldsymbol{w}_\ell = \boldsymbol{w}_1^i$. Define (46) as

$$\mathcal{T}_\ell = \begin{cases} [\prod_{i=\ell+1}^{\mathcal{L}-1} \mathcal{Q}_i(\boldsymbol{\mathcal{D}})] \times d_{x_j} \mathcal{Q}_\ell(\boldsymbol{\mathcal{D}}) \times [\prod_{i=\ell-1}^1 \mathcal{Q}_i(\boldsymbol{\mathcal{D}})], \ell \geq 2 \\ [\prod_{i=\ell+1}^{\mathcal{L}-1} \mathcal{Q}_i(\boldsymbol{\mathcal{D}})] \times d_{x_j} \mathcal{Q}_\ell(\boldsymbol{\mathcal{D}}), \ell = 1 \\ d_{x_j} \mathcal{Q}_\ell(\boldsymbol{\mathcal{D}}) \times [\prod_{i=\ell-1}^1 \mathcal{Q}_i(\boldsymbol{\mathcal{D}})], \ell = \mathcal{L}-1 \end{cases} \quad (46)$$

Finally, the *Hessian* matrix is derived as follows:

$$\frac{\partial^2 \bar{l}_l(t)}{\partial x_i(t) \partial x_j(t)} = \boldsymbol{w}_L \left[ \sum_{\ell=1}^{\mathcal{L}-1} \mathcal{T}_\ell \right] \quad (47)$$

## V. NUMERICAL RESULTS

The performance of the proposed method is tested in the IEEE 39-bus system and it is divided into the sending and the receiving areas as shown in [17]. The wind farm with rated power 500 MW is connected to bus 17. The cut-in, rated and cut-out wind speeds are 5.2 m/s, 11.5 m/s, and 25 m/s, respectively. The cost of wind generation curtailment is 500 $/MW. One ESS is also connected at bus 17, whose rated capacity is 100 MWh with lower and upper bounds of 10 MWh and 100 MW. We assume that the wind farm has adequate reactive reserves and low voltage ride-through capability. For time-domain simulations, the period is 3s with time step 0.05s. The studied contingencies are given in Table I.

TABLE I
CONSIDERED CONTINGENCIES FOR IEEE 39-BUS SYSTEM

| Line | 1-39 | 2-3 | 18-3 | 16-15 |
|---|---|---|---|---|
| Type | 3-phase | 3-phase | 3-phase | 3-phase |
| Notation. | C1 | C2 | C3 | C4 |

### A. TTC Surrogate Model Construction

The regularization parameters of different machine learning algorithms of TTC surrogate model constructions are shown as follows: 1) for the Elastic Net: $\lambda_1^E = 0.5$, $\lambda_2^E = 0.5$; 2) for SLNN and DLNN with a different number of layers, the settings are in Table II. For {40-20} in Table II, it means that the number of neurons from the first hidden layer to the last hidden layer is 40 and 20, respectively. Using (13) and (14) to sample operation conditions and (10)-(12) to compute TTC of each sampled condition, 15,000 samples are generated, in which 85% of samples are for training while the rest 15% is used for testing. The MSE for EN is 0.0553 pu, while those for SLNN, 2-layer DLNN, 3-layer DLNN and 5-layer DLNN are 0.0173 pu, 0.0040 pu, 0.0027 pu and 0.0023 pu, respectively. It can be found that multi-layer DLNN outperforms EN and SLNN and is used in the subsequent tests.

TABLE II
PERFORMANCE OF EACH TTC SURROGATE MODEL ON TEST SETS

| Models | SLNN | 2-layer DLNN | 3-layer DLNN | 5-layer DLNN |
|---|---|---|---|---|
| $\gamma$ | 0.5 | 0.5 | 0.5 | 0.5 |
| Layer Stru. | 10 | {40-20} | {80-40-20} | {40-20-10-5-2} |
| Acti. Func. | *Sigmoid* | *Sigmoid* | *Sigmoid* | *Softplus* |
| MSE/ p.u. | 0.0173 | 0.0040 | 0.0027 | 0.0023 |

TABLE III
DIFFERENT SURROGATE-ASSISTED MODELS FOR OPERATIONAL CONTROL

| Models | without TTC constraints | static TTC constraints | EN-assisted | SLNN-assisted | *Sigmoid* DLNN-assisted | *Softplus* DLNN-assisted |
|---|---|---|---|---|---|---|
| Notation | M0 | M-*S* | M1 | M2 | M3-*L* | M4-*L* |

where $L$ denotes the number of hidden layers of DLNN

### B. Rolling Time-Horizon Operational Control

The performance of the proposed method for operational control is assessed in this section. Specifically, the optimization period is 2~25 *h*, the length of the rolling time-horizon is 24 h, and the dispatch time step is 1-hour considering load and generation forecasting uncertainties. Different surrogate-assisted models for operational control are compared and their descriptions are shown in Table III. Fig. 2 shows the actual TTC, the estimated TTC, the power flow, and the absolute error profiles of tie-line power flow 18-3 using methods M0, M-*S*, M1, M2, M3-2, M3-3, and M4-5. Based on the results in Fig. 2, the post-fault transient trajectories of rotor angle differences after operational controls at time=14*h* are shown in Fig. 3.

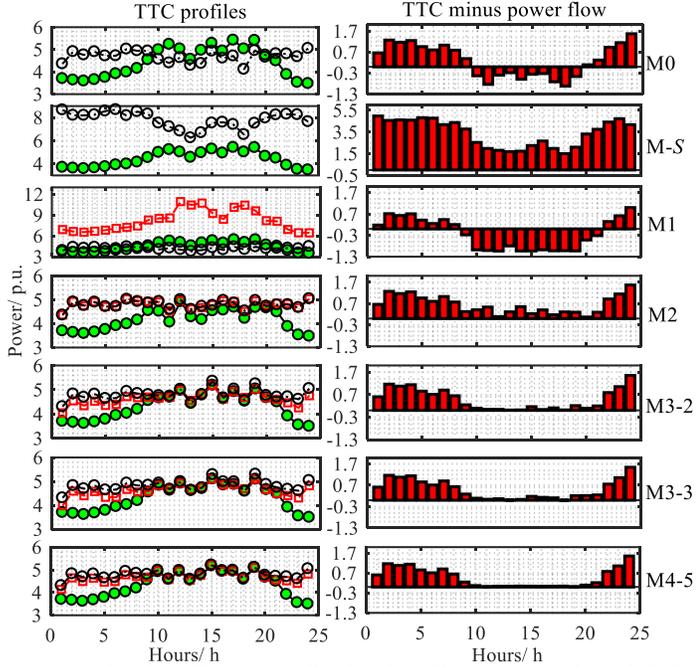

(a) The operation conditions of tie-line 1-39 by several SA-TCOP models and the histogram of TTC minus power flow

⊶- Actual TTC profile  ⊟- Predicted TTC profile  ⊖- Power flow of Tie-line

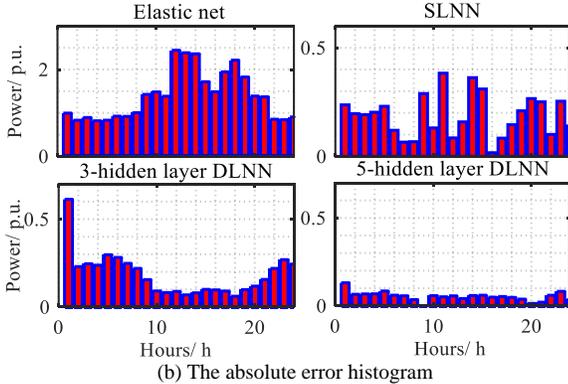

(b) The absolute error histogram

Fig. 2 The conditions of tie-line 1-39 after operational control and the absolute error between actual TTC profiles and TTC estimations.

From Fig. 2, we can find that for M0 and M1, their power flows of the tie-line 1-39 are larger than their TTC during the period 10-19. This means that both M0 and M1 are unable to fulfill the secure planning requirement. Note that M0 does not consider the security constraints and this leads to incorrect operational controls. Although M1 has considered the security constraints, the EN-assisted model cannot accurately follow the TTC constraints and the system loses stability even after operational control, see Fig. 3 for example. By contrast, with a more reliable NNs-based surrogate model, see M2 and M3-3, the system can be controlled effectively without suffering instability issues. It is interesting to observe from Fig. 2 and Fig. 3 that using M-*S*, despite the TTC is much larger than the power flow over the whole optimization period, transient instability still occurs under the contingency C1. Thus, the consideration of transient stability constraints in the TTC model is mandatory to prevent the system from instability in the presence of severe faults. Another observation is that, although all three machine learning algorithms perform well on test sets, their generalization ability cannot always be ensured, see Table II and Fig. 2(b). Among them, DLNNs show the best generalization ability, and this means that powerful NNs should be used when building the TTC surrogate model.

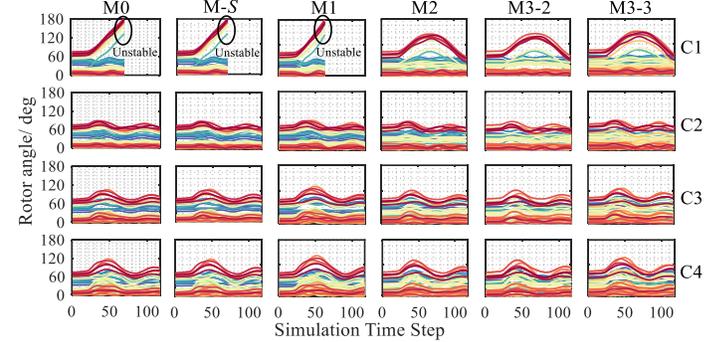

Fig. 3 The post-fault transient trajectories of rotor angle differences using different surrogate model-based operational controls at time=14*h*.

TABLE IV
THE COSTS OF OPTIMAL GENERATION BY DIFFERENT MODELS

| Model. | M0 | M-*S* | M1 | M2 | M3-2 | M3-3 | M4-5 |
|---|---|---|---|---|---|---|---|
| Total Cost/ $ | 56230 | 56229 | 56237 | 56552 | 57211 | 56742 | 56509 |

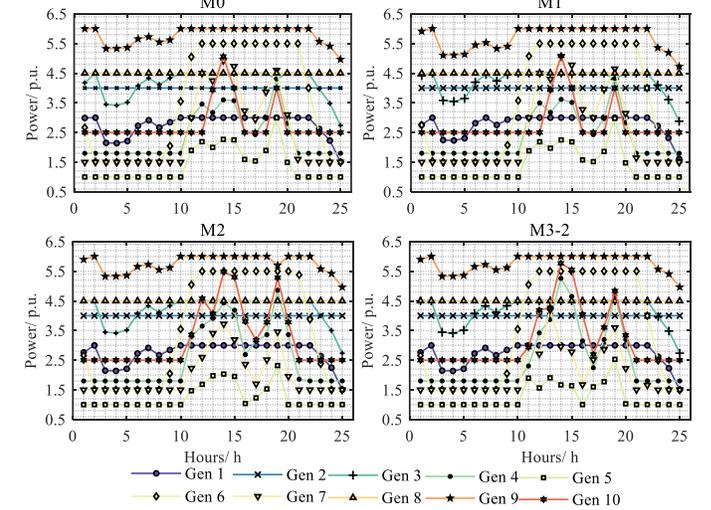

(a) The generation profiles over the optimization horizon
The output of generators at 14h

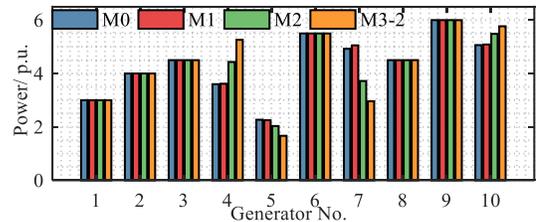

(b) The output of generators at 14h
Fig. 4 The outputs of generators by different models.

The total cost during hours 2-25 period is shown in Table IV. It is found that EN based TTC renders the TTC constraints in the M1 to be inactive. This leads to similar generation dispatch costs for M1 and M0. Note that complying with TTC constraints will increase costs but would ensure system security. Among the models with valid TTC constraints (M2, M3-2, M3-3, and M4-5), M3-2 has the largest surrogate modeling error (MSE: 0.3936 p.u.) and therefore yields the highest costs. MSEs of M2, M3-3, and M4-5 are respectively 0.0440 p.u., 0.1359 p.u., and 0.0118 p.u. The costs for M2, M3-3 and M4-5 decrease by \$659, \$469 and \$702 as compared to M2. Thus, it can be concluded that a better surrogate model contributes to more economical operation strategies.

To further analyze the surrogate model performance in detail, the optimal power generation profiles are given in Fig. 4. Comparing the generation curves optimized by 4 different surrogate model-assisted methods during hour 14, we can observe that there are variations $\Delta x_{\text{Gen}} = [1.075, -0.613, -1.952, 1.557]$ (p.u.) on the outputs of generators 4, 5, 7 and 10 from methods M0 to M3-2. Based on $\nabla(\eta_{18\text{-}3})|_{x_0}^{\text{2-DLNN}}$ calculated via the rules of M3-2, we can roughly compute the variance on the constraint $\eta_{18\text{-}3}$ using $\Delta x_{\text{Gen}}$, which equals 0.086 p.u and this is quite dangerous to system security. The above analysis indicates that $\eta_{18\text{-}3}$ is not just affected by $\Delta x_{\text{Gen}}$, but also by other optimization variables (i.e., $\Delta x$) in the system. Accordingly, $\Delta\eta_{18\text{-}3}$ from M0 to M3-2 should be estimated through $\Delta x \cdot \nabla(\eta_{18\text{-}3})|_{x_0}^{\text{2-DLNN}} \cdot \nabla(\eta_{Tie})|_{x_0}^{\text{2-DLNN}}, Tie \in \{18\text{-}3, 16\text{-}15\}$, and their results are shown in Fig. 5.

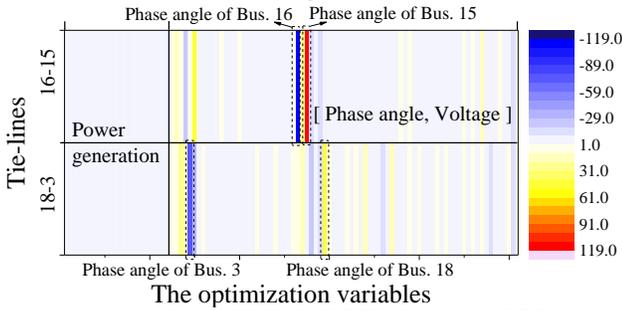

Fig. 5 The *Jacobian* matrices of $\nabla(\eta_{18\text{-}3})|_{x_0}^{\text{2-DLNN}}$ and $\nabla(\eta_{16\text{-}15})|_{x_0}^{\text{2-DLNN}}$ for M0 calculated using 2-layer DLNNs.

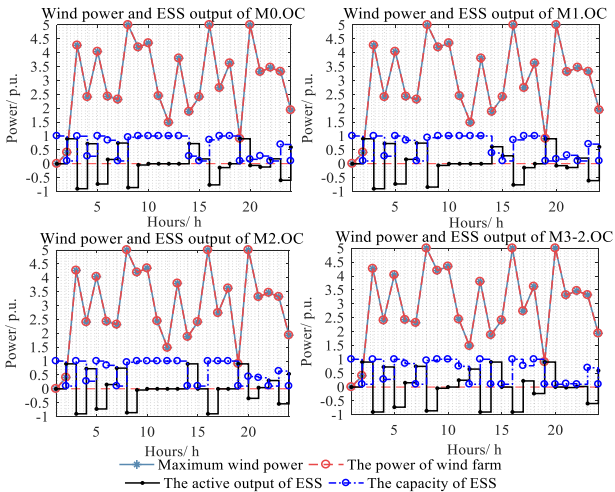

Fig. 6 The optimal output of ESS and wind farm by different models.

To demonstrate the DLNNs-based TTC surrogate model, the most influential variables against TTC should be determined first. The optimization variables are ranked in terms of $\left|\nabla(\eta_{18\text{-}3})|_{x_0}^{\text{2-DLNN}}\right|$ in descending order, and the first two variables are plotted in Fig. 5, i.e., the phase angles at bus 3 ($\delta_3$) and bus 18 ($\delta_{18}$). According to the physical system model, the two variables that have the largest impacts on the active flow of tie-line 18-3 should be $\delta_3$ and $\delta_{18}$. The gradient information obtained by our method is consistent with that conclusion.

The effect of ESS on the TTC security-constrained model is also assessed here. The optimal generation profiles of the ESS and the wind farm for different models are shown in Fig. 6. It can be found that all models enable full utilization of wind energy. In comparison with the load curve, we find that the wind power operates in an inverse load peak regulation during the period 18-19 h. The load consumption increases from 4177.24 MW to 4433.67 MW, whereas the wind power drops from 362.01 MW to 90.31 MW. In this case, the generators are required to balance a total of 528.18 MW, which is the maximum power gap over the optimization period. Therefore, ESS reaches the maximum active output in hour 19. Note that there are only slight differences among all the optimal ESS output curves. Hence, a small-sized ESS is mainly used for peak load shifting and therefore yields a minor effect on TTC.

### C. Comparison with Sensitivity-based Method

To compare the proposed data-driven assisted method with the traditional full model-based optimization approach, the corrective control model (CCM) method is used. The purpose of CCM is to tune the control variables of M0 to stabilize the post-fault system, which can be formulated as follows:

$$\min \sum_{t \in \mathbb{U}} \Delta t \sum_{i \in \mathbb{G}} C_i \left(P_{i,0} + \Delta x_i\right)$$

$$s.t. \quad (3)\text{-}(7)$$

$$\eta_{i,k-1} + \Delta x_k * \frac{\partial \eta_{i,k-1}}{\partial x_{k-1}}|_{x_0 = \text{M0}} < 0, i \in \mathbb{K}$$

where $\Delta x$ is the corrective variables used for adjusting the control variables; $\eta_{i,k}$ is the security margin at the $k$-th iteration; $\Delta x_k$ is the corrective variables at the $k$-th iteration; $\frac{\partial \eta_{i,k}}{\partial x_k}$ is the sensitivity of security margin to the changes in control variables at the $k$-th iteration and it can be roughly estimated by numerical simulations. $\eta_{i,0}$ denotes the security margin of M0; $x_0$ represents the initial conditions, and $x_0 = $ M0 implies that CCM is carried out on the baseline operation determined by M0. The results after the corrective re-dispatch are shown in Fig. 7. It can be observed that CCM enables secure operations. However, the generation cost by CCM is \$58262, which is much larger than the costs by M2, M3-2, M3-3, and M4-5 as given in Table IV. Moreover, CCM takes around 20 minutes for sensitivity calculation in every iteration that is much slower as compared to our proposed method shown in Table VIII in the next section. This means that CCM is very time-consuming and hard to be used for large-scale power systems.

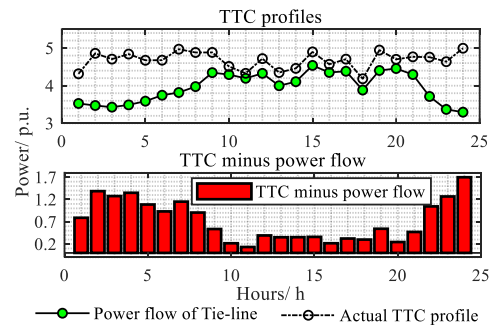

Fig. 7 The tie-line power flow 1-39 after CCM control and the absolute error between actual TTC profiles and TTC estimations.

### D. Computational Efficiency Assessment

Table VIII shows the computing times of all methods. All statistics are taken from the mean value of rolling optimizations. A shorter horizon results in less computing time. For instance, when a 6-hour length of the horizon is chosen, the solving times of the models M3-2, M3-3, and M4-5 can be reduced to around 10 seconds. Note that high-performance computing techniques, e.g., distributed or parallel computing, GPU acceleration, etc., could be further applied for a faster SA-TCOP. Several heuristic methods [10]–[12], [21] have been adopted to serve as comparative studies as well. All these methods take around 1 day to solve the proposed TCOP problem as they suffer from "curse of dimensionality" when solving the TCOP containing 2448 variables and 2928 constraints.

TABLE VIII
THE COMPUTATIONAL COSTS BY DIFFERENT TESTED MODELS

| Model/ Surrogate | M0 | M-$S$ | M1 | M2 | M3-2 | M3-3 | M4-5 |
|---|---|---|---|---|---|---|---|
| Iterative times | 32.7 | 33.3 | 60.1 | 74.1 | 73.9 | 72.3 | 70.8 |
| Mean Sol. time /s | 74.6 | 82.8 | 127.7 | 138.7 | 135.5 | 136.9 | 135.4 |

## VI. CONCLUSION

This paper proposes a surrogate model-based method for TTC constrained operation planning (TCOP) of power systems considering wind generators and ESSs. The essential idea is to develop a computationally cheap surrogate model to replace the original complicated and time-consuming constraints, i.e., the TTC. This is achieved via the deep learning algorithms. However, the deep learning-based surrogate model further involves the "black-box" type rules that are challenging for existing optimization methods. To deal with that, we derive the analytical Jacobian and Hessian matrices of the implicit surrogate models. This allows us to transform the "black-box" type rules into an analytical formulation that can be easily solved by the interior point method. Simulation results show that the proposed method enables economic and efficient operational control of the system. It achieves much higher computational efficiency than other alternatives.